\documentclass[journal]{IEEEtran}

\usepackage{color}

\usepackage{cite}

\usepackage{textcomp}

\usepackage{scalerel}
\usepackage{tikz}
\usetikzlibrary{svg.path}

\definecolor{orcidlogocol}{HTML}{A6CE39}
\tikzset{
	orcidlogo/.pic={
		\fill[orcidlogocol] svg{M256,128c0,70.7-57.3,128-128,128C57.3,256,0,198.7,0,128C0,57.3,57.3,0,128,0C198.7,0,256,57.3,256,128z};
		\fill[white] svg{M86.3,186.2H70.9V79.1h15.4v48.4V186.2z}
		svg{M108.9,79.1h41.6c39.6,0,57,28.3,57,53.6c0,27.5-21.5,53.6-56.8,53.6h-41.8V79.1z M124.3,172.4h24.5c34.9,0,42.9-26.5,42.9-39.7c0-21.5-13.7-39.7-43.7-39.7h-23.7V172.4z}
		svg{M88.7,56.8c0,5.5-4.5,10.1-10.1,10.1c-5.6,0-10.1-4.6-10.1-10.1c0-5.6,4.5-10.1,10.1-10.1C84.2,46.7,88.7,51.3,88.7,56.8z};
	}
}

\newcommand\orcidicon[1]{\href{https://orcid.org/#1}{\mbox{\scalerel*{
				\begin{tikzpicture}[yscale=-1,transform shape]
				\pic{orcidlogo};
				\end{tikzpicture}
			}{|}}}}

\usepackage{hyperref}

\ifCLASSINFOpdf
\else
\fi

\usepackage{amsmath}

\usepackage{graphicx}

\usepackage{array}

\usepackage[normalem]{ulem}
\usepackage{xcolor}
\usepackage[utf8]{inputenc}
\usepackage[T1]{fontenc}

\usepackage{units}

\begin{document}
\begin{titlepage}

{\onecolumn \copyright~2021 IEEE. Personal use of this material is permitted. Permission from IEEE must be obtained for all other uses, in any current or future media, including reprinting/republishing this material for advertising or promotional purposes, creating new collective works, for resale or redistribution to servers or lists, or reuse of any copyrighted component of this work in other works.

A. P. Jovanović, M. N. Stankov, D. Loffhagen, and M. M. Becker, "Automated Fluid Model Generation and Numerical Analysis of Dielectric Barrier Discharges Using Comsol," \textit{IEEE Trans.  Plasma Sci.}, vol. 49, pp. 3710-3718, 2021, doi: 10.1109/TPS.2021.3120507.\\ URL: https://ieeexplore.ieee.org/document/9600812
}
\end{titlepage}

\title{Automated fluid model generation and numerical analysis of 
dielectric barrier discharges using Comsol}

\author{{Aleksandar P. Jovanović \orcidicon{0000-0002-7104-6466}, Marjan N. Stankov \orcidicon{0000-0002-6132-2754}, Detlef Loffhagen \orcidicon{0000-0002-3798-0773}, Markus M. Becker \orcidicon{0000-0001-9324-3236}}%
\thanks{The authors are with the  
Leibniz Institute for Plasma Science and Technology (INP), Felix-Hausdorff-Straße 2, 17489 Greifswald, Germany}%
}

\maketitle

\begin{abstract}
MCPlas is introduced as a powerful tool for automated fluid model generation with application to the analysis of dielectric barrier 
discharges operating  in different regimes. MCPlas consists of a number of MATLAB\textsuperscript{\textregistered} scripts and uses the  COMSOL Multiphysics\textsuperscript{\textregistered} module LiveLink\textsuperscript{\texttrademark} for MATLAB\textsuperscript{\textregistered} to 
build up 
equation-based COMSOL Multiphysics\textsuperscript{\textregistered} models from scratch. 
The present contribution highlights how MCPlas is used to implement time-dependent models for non-thermal plasmas in spatially one-dimensional and axisymmetric two-dimensional geometries and stresses out the benefit of automation of the modelling procedure. The modelling 
codes  generated by MCPlas  
are  used to  study  diffuse  and  filamentary dielectric barrier discharges  in  argon at sub-atmospheric and atmospheric pressure, respectively. The  seamless  transition between different levels of model complexity with respect to the
considered model geometry is demonstrated. The presented investigation of a single-filament  dielectric  barrier  discharge  interacting  with a  dielectric  surface  shows that  complex phenomena of high technological relevance can be tackled by using plasma models implemented in COMSOL Multiphysics\textsuperscript{\textregistered} via MCPlas.
\end{abstract}


\IEEEpeerreviewmaketitle

\section{Introduction}
\IEEEPARstart{D}IELECTRIC barrier discharges (DBDs) are frequently used to generate non-thermal plasmas for technological applications, such as ozone generation, surface processing and plasma medicine~\cite{Kogoma-1994-ID5471, Fridman-2005-ID2283, Kogelschatz-1997-ID2958, Kogelschatz-2003-ID2019, Brandenburg-2017-ID4908}. DBDs are characterised by the presence of a dielectric layer on at least one electrode, which limits the electric current and prevents transition to an arc plasma. 
Depending on the conditions, DBDs can be diffuse or filamentary, in some cases self-organised patterns occur~\cite{Kogelschatz-2002-ID3556}. Due to their broad use, it is of crucial importance to get a detailed understanding of the physical and chemical processes occurring in DBDs. 
A common way to do this is by applying fluid-Poisson models, which couple fluid equations for the plasma species with Poisson's equation for the electric potential~\cite{Braun-1992-ID749, Steinle-1999-ID1363,  Gibalov-2000-ID1450, Papageorghiou-2009-ID2548, Gibalov-2012-ID2757, Boeuf-2013-ID3452, Becker-2013-ID3200}. 
The computational efficiency of these models allows one to take into account complex physical and chemical processes occurring during the plasma production, which makes them suitable for practical applications. 

In the present manuscript, a toolbox combining functionalities 
of 
COMSOL Multiphysics\textsuperscript{\textregistered}~\cite{ComsolMultiphysics} (in further text Comsol) 
and MATLAB\textsuperscript{\textregistered}~\cite{MATLAB_R2018b} (in further text Matlab), named MCPlas,  
is introduced for the automated implementation of fluid-Poisson models in Comsol.
MCPlas consists of a number of 
Matlab  scripts using the functionality provided by the COMSOL module LiveLink\textsuperscript{\texttrademark} for MATLAB\textsuperscript{\textregistered} 
(in further text Livelink module) 
to build up the plasma models in Comsol. 
This replaces the manual definition of equations, which becomes a tedious and error-prone process for complex models often taking into account tens to hundreds of species and hundreds to thousands of reaction kinetic processes~\cite{Simek-2018-ID5457,Ponduri-2016-ID3865,DeBie-2015-ID3918,Lazarou-2016-ID4072, Loffhagen-2021-ID5481}. Therefore, the main aim of this work is to stress out %
the benefit of automation and to illustrate how this can be achieved by using MCPlas.

In order to illustrate the use of MCPlas for automated code implementation, two test cases are presented: $1$) a diffuse DBD in argon at sub-atmospheric pressure, and $2$) a filamentary atmospheric-pressure DBD in argon. The numerical results obtained for these application cases are analysed in detail to explain the main features of the different discharges. The present study substantially extends the initial results presented in~\cite{Jovanovic_CC2020} 
and 
demonstrates 
the wide-range applicability of MCPlas. An earlier, simplified version of MCPlas was applied in the benchmark simulation study on positive streamers in air \cite{Bagheri-2018-ID5240}.

The paper is structured as follows. Section \ref{sec:mcplas} introduces the MCPlas toolbox by illustrating the structure of the code and describing its main features. 
The results and discussion section \ref{sec:results} describes the fluid-Poisson model implementation obtained from MCPlas and introduces the two case studies as exemplary applications.
Finally, main conclusions are given in section~\ref{sec:conclusion}.

\section{MCPlas toolbox}
\label{sec:mcplas}
The Matlab-Comsol toolbox for plasma modelling was developed for automated implementation of fluid-Poisson models in Comsol. It consists of a number of Matlab scripts and uses the Livelink module to set up the system of partial differential equations using equation-based modelling. This means that all equations are implemented on the basis of the basic differential equation interfaces in {Comsol} without using additional modules. 

\begin{figure}[htb]%
	\centering%
	\includegraphics[width=80mm]{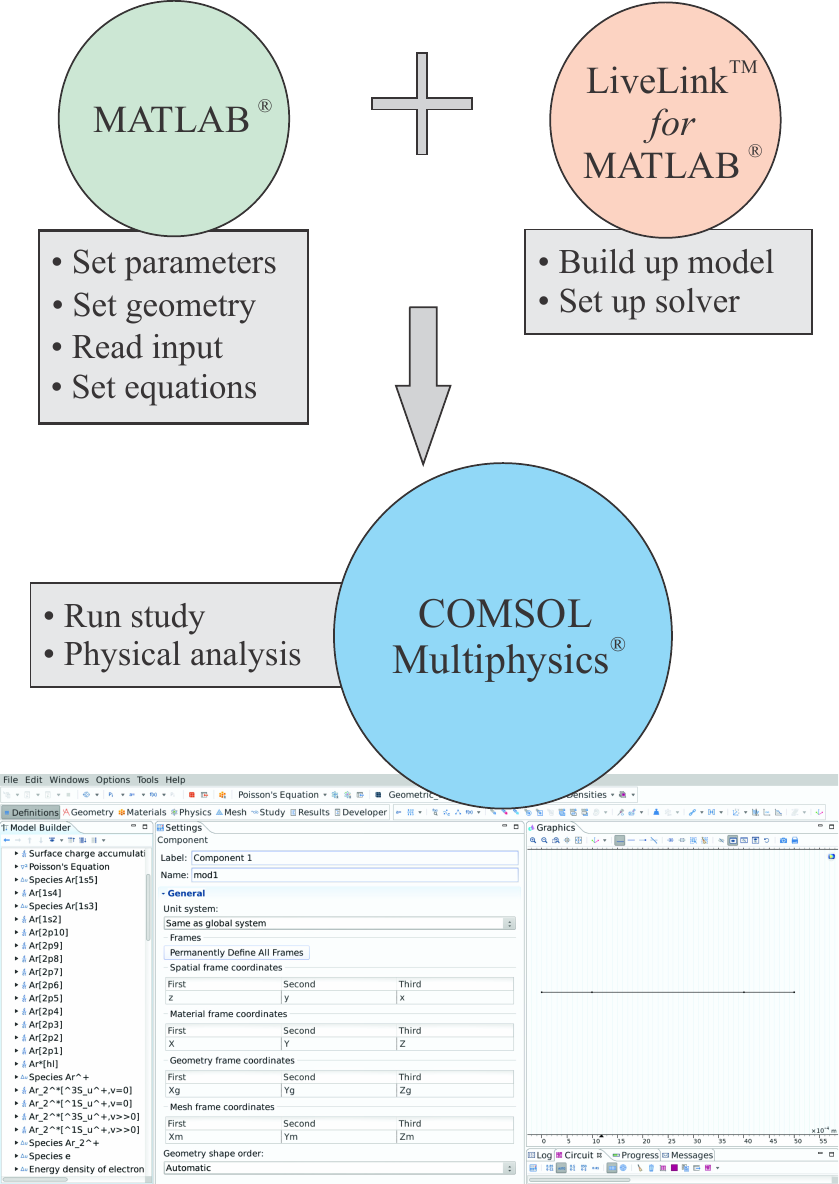}%
	\caption{Model generation and analysis workflow of MCPlas.}%
	\label{fig:workflow}%
\end{figure}

The workflow for the implementation and usage of a plasma model in {Comsol} using MCPlas is visualised in Fig.~\ref{fig:workflow}. 
The first step in the Matlab code sets up the relevant parameters. This includes information about the discharge geometry, specific domains (plasma, dielectric etc.), boundary  selections (e.g. powered electrode, grounded electrode, dielectric-plasma interface), and modelling procedure. Depending on these parameters, the geometry is defined and a proper form of coordinates (Cartesian or cylindrical), boundary system, and the equations are implemented.

The input data consists of a list of the plasma species considered, their properties, and plasma-chemical reactions together with rate and transport coefficients. This input data is read in next. 
It is important to note that the implementation of all balance 
equations with corresponding reaction kinetic processes is automated by using 
the species list and reaction kinetic scheme prepared in advance. 
These input files specify which species, collision processes and radiation processes are taken into account, and directly determine the balance equations to be solved. 
The input data also specify the set of physically motivated boundary conditions and initial values  to complete the coupled system of partial differential equations.

After these steps, the {Livelink} module is employed to 
build up a {Comsol} model by setting up the definitions of physical quantities, coefficients and governing equations with their initial and boundary conditions.
Additional input variables are used to set up the numerical solver for efficient solution of the nonlinear coupled system of equations. 
Note that MCPlas is able to automatically generate plasma models in Comsol involving an arbitrary number of species and reaction kinetic processes depending on the general input files defined by the user. 
While the general set of equations and boundary conditions are pre-defined by the MCPlas code, the gas (mixture), number of considered species, reaction kinetic scheme, geometry, and external discharge parameters can be changed by the user via the input parameters and input files. 

The last step in the modelling process is to run the study, which can either be done by means of the Comsol user interface or by starting the solution procedure directly from Matlab using the Livelink module and extending MCPlas accordingly. The Comsol user interface is used here to perform the analysis (cf. Fig.~\ref{fig:workflow}).

\section{Results and Discussion}
\label{sec:results}
The application of MCPlas 
results in a system of fluid equations coupled with Poisson’s equation as typically used for the numerical analysis of non-thermal plasmas for which a hydrodynamic description is valid.
The following subsections first present and discuss the important features of the automated model implementation provided by MCPlas. Second, they illustrate the robustness and applicability of the model provided by MCPlas for the analysis of DBDs using two test cases.

\subsection{Automation of model implementation}
\label{sec:model}
\subsubsection{Governing equations}
The automatically generated Comsol model comprises
balance equations for the particle number densities $n_p$ of the plasma species $p$ (electrons, various ions and neutral particles in their ground and excited states) specified by the input files
        	
\begin{equation}
\label{ContEq}
\frac{\partial n_p}{\partial t} + \nabla \cdot \mathbf{\Gamma}_p=S_{p}\,,
\end{equation}

\noindent the electron energy balance equation for the energy density of electrons $w_\mathrm{e}$
               	
\begin{equation}
\label{EEBEq}
\frac{\partial w_\mathrm{e}}{\partial t} + \nabla \cdot \mathbf{Q}_\mathrm{e} = -e_0 \mathbf{E} \cdot \mathbf{\Gamma}_\mathrm{e} +\widetilde{S}_\mathrm{e}\,,
\end{equation}

\noindent and Poisson’s equation for the electric potential $\phi$

\begin{equation}
\label{PoissonEq}
      	-\varepsilon_0 \varepsilon_\mathrm{r} \nabla^2 \phi = \sum_p q_{p} n_{p}\,.
\end{equation}  	

\noindent Here, $\mathbf{\Gamma}_p$ and $\mathbf{Q}_\mathrm{e}$ are the particle fluxes and the electron energy flux, $S_p$ and $\widetilde{S}_\mathrm{e}$ are source terms describing the gain and loss of particles and electron energy 
due to
reaction kinetic processes, $e_0$ denotes the elementary charge, $\mathbf{E} = - \nabla\phi$ is the electric field, $\varepsilon_0$ and $\varepsilon_\mathrm{r}$ represent the vacuum permittivity and relative permittivity of the medium, respectively, and $q_p$ is the charge of species $p$. For the fluxes of particles and the 
electron 
energy flux, the drift-diffusion approximations 
\begin{align}
\label{Fluxes}
\mathbf{\Gamma}_p &= \textrm{sgn}(q_p)\,b_p\,\mathbf{E}\, n_p - \nabla (D_p \, n_p)\,,\\
\mathbf{Q}_\mathrm{e} &= -\,\widetilde{b}_\mathrm{e}\,\mathbf{E}\, w_\mathrm{e} - \nabla (\widetilde{D}_\mathrm{e} \, w_\mathrm{e})
\end{align}
are used, where $b_p$ and $D_p$ are the mobility and diffusion coefficients for species $p$, and $\widetilde{b}_\mathrm{e}$ and $\widetilde{D}_\mathrm{e}$ are the mobility and diffusion coefficient for energy transport of electrons, respectively. 

Note that the basic set of differential equations can be changed by modifying the MCPlas programme code, e.g. to take into account an equation and additional terms for the gas flow velocity or to use 
different kinds of approximations for the fluxes.  
It should be also 
mentioned that MCPlas provides an option to solve these balance  
equations in logarithmic representation, where the equations are solved for the logarithm of the particle number density and the electron energy density. This prevents large scale variations and possible negative values in the solution.

\subsubsection{Boundary and initial conditions}

For Poisson’s equation, Dirichlet boundary conditions are set, i.e. the potential is set to $\phi = 0$ at the grounded electrode, and $\phi=U_\mathrm{a}$  at the powered electrode. 
Here, $U_\mathrm{a}$ denotes the applied voltage, which can be constant or given as a time-dependent function describing the temporal evolution of the voltage signal.

For modelling of DBDs, the additional interface condition $-\varepsilon_0 \varepsilon_\mathrm{r} \boldsymbol{E} \cdot \boldsymbol{\nu}=\sigma$ is used at the plasma-dielectric interface to account for the change of the electric field due to accumulated surface charges. 
In this case, the balance equation 
\begin{equation}
\frac{\partial \sigma}{\partial t} = \sum_p q_p\mathbf{\Gamma}_p \cdot \boldsymbol{\nu}
\label{eq:DBDSurfaceBoundary}
\end{equation}
for the surface charge density $\sigma$ accumulated at the dielectric boundaries facing to the plasma volume is added to the model, where $\boldsymbol{\nu}$ is the outward normal vector to the respective boundary surface \cite{Becker-2013-ID3200}.

For the particle balance equations, boundary conditions according to Hagelaar \textit{et al.}~\cite{Hagelaar-2000-ID1480} are applied

\begin{enumerate}

\item for the  
balance equation (1) of heavy particles

\begin{equation}
\boldsymbol{\nu}  \cdot  \mathbf{\Gamma}_{p} = \frac{1-r_p}{1+r_p} \Big( | b_p \mathbf{E} n_p| + \frac{1}{2} v_{\mathrm{th}, p} n_p \Big)\,;
\label{eq:ContBC}
\end{equation}

\item for the  
balance equation (1) of electrons

\begin{align}
\boldsymbol{\nu}  \cdot  \mathbf{\Gamma}_\mathrm{e} &= \frac{1-r_\mathrm{e}}{1+r_\mathrm{e}} \Big( |b_\mathrm{e} \mathbf{E} n_\mathrm{e}| + \frac{1}{2} v_\mathrm{th, e} n_\mathrm{e} \Big) \nonumber \\
&\quad- \frac{2}{1+r_\mathrm{e}} \gamma \sum_{i} \textnormal{max}(\mathbf{\Gamma}_i\cdot\boldsymbol{\nu}, 0)\,;
\end{align}

\item for the electron energy balance equation (2)
 
\begin{align}
\boldsymbol{\nu} \cdot \mathbf{Q}_\mathrm{e} &= \frac{1-r_\mathrm{e}}{1+r_\mathrm{e}} \Big( |\widetilde{b}_\mathrm{e} \mathbf{E} w_\mathrm{e}| + \frac{1}{2} \widetilde{v}_\mathrm{th, e} w_\mathrm{e} \Big) \nonumber \\
&\quad- \frac{2}{1+r_\mathrm{e}} u_\mathrm{e}^\gamma \gamma \sum_{i} \textnormal{max}(\mathbf{\Gamma}_i \cdot \boldsymbol{\nu}, 0)\,.
\label{EEBE_bc}
\end{align}   	

\end{enumerate}

\noindent Here, $v_{\mathrm{th},p}$ is the thermal velocity of the respective species, $\widetilde{v}_\mathrm{th,e}=2k_\mathrm{B} T_\mathrm{e} v_\mathrm{th,e}$ with 
the Boltzmann constant $k_\mathrm{B}$ and 
electron 
temperature $T_\mathrm{e} = 2 w_\mathrm{e}/(3 k_\mathrm{B} n_\mathrm{e})$, 
$r_{p}$ are the reflection coefficients of species $p$, $\gamma$ is the secondary electron emission coefficient, 
and 
$u_\mathrm{e}^{\gamma}$ denotes the mean energy of secondary electrons. 
It should be  
noted that other sets of boundary conditions can be implemented as well by modifying the corresponding module of the MCPlas toolbox. 

As initial values, quasi-neutral conditions with uniform number densities for all particle species  
are used. Similarly, the 
initial electron energy density is assumed to be uniform. 

\subsubsection{Source terms}
The  source terms $S_p$ and $\widetilde{S}_\mathrm{e}$ are defined as 
\begin{equation}
\label{ContEqSourceTerm}
S_p = \sum_{j=1}^{N_\mathrm{r}} (G_{pj} - L_{pj}) R_j\,,
\end{equation}
\begin{equation}
\label{EEBEqSourceTerm}
\widetilde{S}_\mathrm{e} = \sum_{j=1}^{N_\mathrm{r}} \Delta \varepsilon_j R_j\,,
\end{equation} 
where $R_j$ is the reaction rate
\begin{equation}
R_j = k_j \prod_{i=1}^{N_\mathrm{s}} n_i^{\beta_{ij}}\,.
\end{equation}
Here, the parameter $\beta_{ij}$ is the partial reaction order of species $i$ in reaction $j$, $k_j$ is the reaction rate coefficient for reaction $j$, $\Delta \varepsilon_j$ represents the electron energy gained or lost in 
reaction $j$, and $N_\mathrm{r}$ and $N_\mathrm{s}$ are the number of reactions and species, respectively.
The gain ($G_{pj}$) and loss  ($L_{pj}$) matrix elements in equation \eqref{ContEqSourceTerm} are defined by the stoichiometric  coefficients  for  the given 
species and reactions.
MCPlas automatically generates these matrices from the species list and reaction scheme as described in section~\ref{sec:mcplas}.
With this, the transition between models of different complexity in terms of the number of considered species and reactions is simplified.
This provides the option, e.g. to easily perform studies on the influence of changes in the considered reaction kinetic model.

\subsubsection{Reaction rate and transport coefficients}

The plasma model implemented by MCPlas involves the rate coefficients for the collision and radiation processes as well as the transport coefficients for all considered species as variable definitions, look-up tables or analytical functions. 
The particular form of the coefficients depends on the given input files specifying the constant value or functional form. 
Various plasma quantities or geometrical parameters can be used as parameters for the function definitions, such as electric field, mean electron energy, gas temperature or characteristic lengths of the system.

A great benefit of this flexibility is that the set of coefficients can be exchanged without additional implementation effort. Thus it is easy, e.g.\ to switch between the local mean energy approximation and the local field approximation for definition of the coefficients of the electron component~\cite{Park-1990-ID656,Hagelaar-2005-ID2276,Grubert-2009-ID2551}.

\subsubsection{Geometry and mesh}
Besides implementation of the various equations with their source terms and coefficients, the application of MCPlas results in a proper definition of the coordinates and boundary system for the given geometry. 
MCPlas supports spatially one-dimensional and two-dimensional models in Cartesian and cylindrical coordinates, so far. The geometry is either created by the Livelink module based on the setup parameters or read from a geometry file prepared in advance.
Furthermore,  MCPlas associates the boundary conditions \eqref{eq:ContBC}--\eqref{EEBE_bc} with the respective boundaries and interfaces of the computational domain.
This is achieved by the definition of domain and boundary selections. 

Once the geometry is set up, the prescribed element sizes for the given domain selections result in an automated generation of the mesh, which might be refined in selected regions.
The possibility of exchanging the geometry for a given plasma model via MCPlas and the automated mesh generation provided by Comsol allows to investigate various setups without additional implementation effort. 

\subsection{Test cases}

Two DBDs in argon at different conditions are used as test cases.  Both test cases use the same number of species and the same reaction kinetic model for the spatiotemporal description of the discharge behaviour in diffuse DBD (first test case) and in a single-filament configuration (second test case), respectively.
The model takes into account electrons plus 22 heavy particle species participating in 409 reaction kinetic processes. A list of the considered argon species with the corresponding energy levels is given in Table~\ref{table_1}. 

\begin{table}[htb]
	\renewcommand{\arraystretch}{1.3}
	\caption{List of heavy particle species considered in the two test cases.}
	\label{table_1}
	\centering
	\begin{tabular}{c c}
		\hline
		Species & Energy level [eV]\\
		\hline
		$\text{Ar[1p}_\text{0}]$ & 0\\
		$\text{Ar[1s}_\text{5}]$ & 11.55\\
		$\text{Ar[1s}_\text{4}]$ & 11.62\\
		$\text{Ar[1s}_\text{3}]$ & 11.72\\
		$\text{Ar[1s}_\text{2}]$ & 11.82\\
		$\text{Ar[2p}_\text{10}]$ & 12.91\\
		$\text{Ar[2p}_\text{9}]$ & 13.08\\
		$\text{Ar[2p}_\text{8}]$ & 13.09\\
		$\text{Ar[2p}_\text{7}]$ & 13.15\\
		$\text{Ar[2p}_\text{6}]$ & 13.17\\
		$\text{Ar[2p}_\text{5}]$ & 13.27\\
		$\text{Ar[2p}_\text{4}]$ & 13.28\\
		$\text{Ar[2p}_\text{3}]$ & 13.30\\
		$\text{Ar[2p}_\text{2}]$ & 13.33\\
		$\text{Ar[2p}_\text{1}]$ & 13.40\\
		$\text{Ar}^\text{*}[\text{hl}]$ & 13.84\\
		$\text{Ar}^+$ & 15.76\\
		$\text{Ar}_2^*[^3\Sigma_\text{u}^+, \text{v}=0]$ & 9.76\\
		$\text{Ar}_2^*[^1\Sigma_\text{u}^+, \text{v}=0]$ & 9.84\\
		$\text{Ar}_2^*[^3\Sigma_\text{u}^+, \text{v}\gg0]$ & 11.37\\
		$\text{Ar}_2^*[^1\Sigma_\text{u}^+, \text{v}\gg0]$ & 11.45\\
		$\text{Ar}_2^+$ & 14.50\\
		\hline
	\end{tabular}
\end{table}

The local-mean-energy approximation \cite{Grubert-2009-ID2551} is applied for the electron component, i.e. the rate coefficients for 
particle and energy transfer processes due to elastic collisions and various inelastic (e.g.\ exciting, de-exciting, ionizing)  collisions of electrons 
as well as the electron transport coefficients are determined in advance 
by solving the  
stationary, spatially homogeneous electron 
Boltzmann equation in multi-term approximation~\cite{Leyh-1998-ID1222}. 
These data are provided in the form of look-up tables as a function of the mean electron energy and used as input files for MCPlas.
Further details about the reaction kinetics and transport properties are given in~\cite{StBeBaWeLo2020PSST125009}.
The reflection coefficients of electrons, ions and excited argon atoms as well as the secondary electron emission coefficient and corresponding energy of secondary electrons 
are taken as in~\cite{Becker-2013-ID3200}. 

Once the input files are prepared, execution of MCPlas  takes only few tens of seconds to build up the Comsol models for the two test cases described in the following subsections.

\subsubsection{Modelling of a diffuse DBD in argon at 500 mbar}
\begin{figure}[htb]
\centering
\includegraphics[width=80mm]{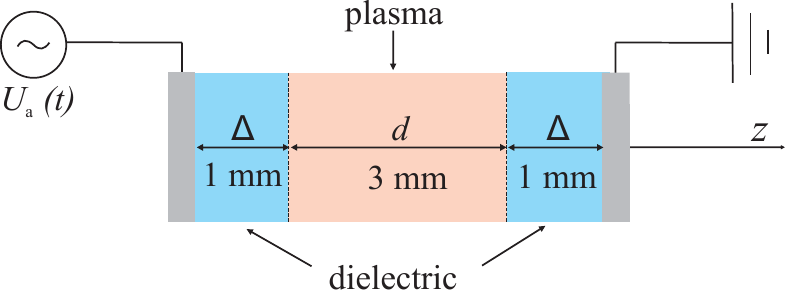}
\caption{Schematic description of the spatially one-dimensional DBD geometry used in test case 1.}
\label{fig_2}
\end{figure}

The first test case for the 
application of MCPlas
is the time-dependent and spatially one-dimensional (1D) modelling of an argon DBD at 500\,mbar in plane-parallel configuration, where both electrodes are covered by glass dielectrics ($\varepsilon_\mathrm{r}=4.2$, $\gamma = 0.02$) as illustrated in Fig.~\ref{fig_2}. 
The model setup is the same as used in Stankov \textit{et al.}~\cite{StBeBaWeLo2020PSST125009}.
The gap between the $1$\,mm  
thick
dielectric layers is $d = 3$\,mm. 
The spatial variation of the plasma in the gap takes place along the $z$-axis between $z = 1$\,mm (left side) and $z = 4$\,mm (right side). 
The diffuse gas discharge is uniform perpendicular to this axis and has an cross-sectional area of $A = 7.26$\,cm$^\text{2}$. The gas temperature is assumed to be constant at 
$T_\mathrm{g} = 300$\,K. 
The sinusoidal voltage $U_\mathrm{a}(t) = U_0 \sin(2 \pi ft)$ is applied at the left electrode, while the right electrode is grounded. The applied voltage has an amplitude of 
$U_0 = 2$\,kV
and a frequency of 
$f = 13.7$\,kHz. 
Hence the period duration is equal to 
$T = 73$\,\textmu s.  
The computational domain is divided into the plasma region with 1500 mesh elements and the dielectric part with 50 elements. 
Initial values of number densities for all species are assumed to be uniform with a number density of $10^{12}$\,m$^\text{-3}$ for heavy particles and $2 \times 10^{12}$\,m$^\text{-3}$ for the electrons. 

Once the Comsol model is set up by using MCPlas, various discretisation methods and parameters can be set to optimise the solution procedure. Here, the finite element method with linear Lagrange elements is used for the balance %
equations of particle number densities and the electron energy balance equation, while quadratic Lagrange elements are applied for the Poisson equation. Discontinuous Lagrange elements of the first order are employed for the balance equation of surface charges. For the temporal discretisation, the backward differentiation formula with adaptive order varying between 1 and 5 is applied. The complete set of equations is  solved in a fully coupled manner and the time-dependent solver utilizes the constant Newton method with damping factor equal to 1. 
The direct solver PARDISO is used for solving the system of linear equations. 
For present calculations the prescribed absolute and relative tolerances are set to  $10^{-4}$. 
It was found that these values are adequate to achieve the required accuracy without hindering an efficient solution of the highly non-linear problem.
Note that the balance equations are solved in logarithmic form to achieve better stability and no additional scaling is used for the solution variables. The calculation was carried out on a 
computer with two Intel\textsuperscript{\textregistered} Xenon\textsuperscript{\textregistered} E5-2667 v2 @ 3.30GHz CPUs and with 125 Gb of RAM available in total. The calculations were carried out on 4 cores and lasted around 1.5 hours per period.

\begin{figure}[!t]
\centering
\includegraphics[width=80mm]{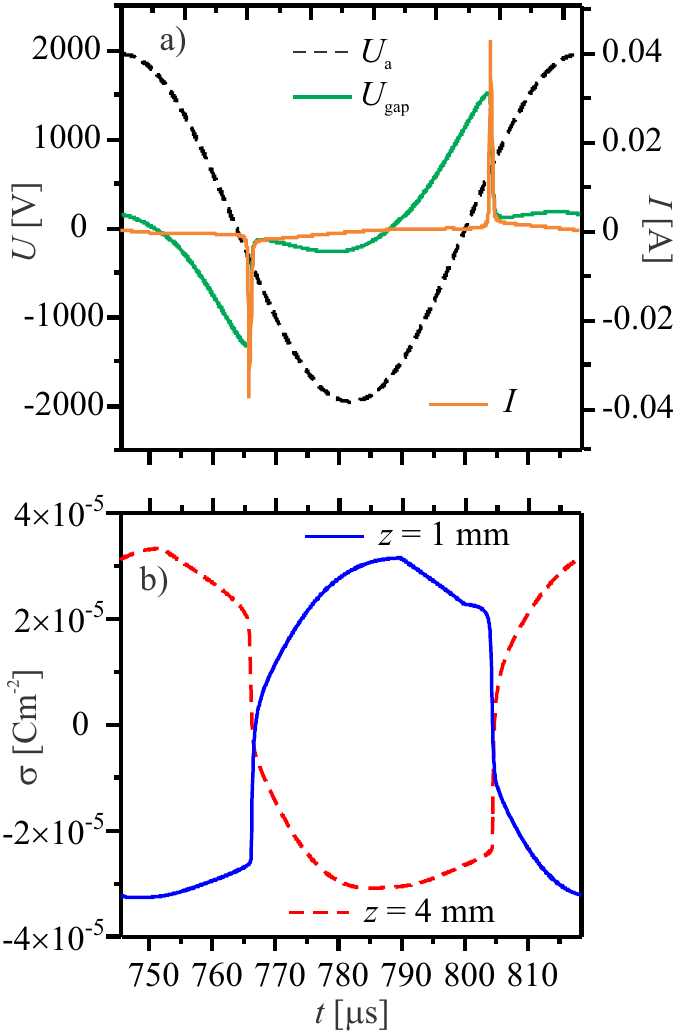}
\caption{Temporal evolution of a) discharge current 
	$I$	and gap voltage $U_\mathrm{gap}$ together with the voltage applied at the powered electrode $U_\mathrm{a}$ and b) surface charge density $\sigma$ on the  dielectrics at $z = 1$\,mm and $z = 4$\,mm  obtained by 
model calculations
	for $p = 500$\,mbar, $U_0 = 2$\,kV and $f = 13.7$\,kHz (test case 1).}
\label{fig_3}
\end{figure}

The calculated discharge current $I$ and gap voltage $U_\mathrm{gap}$ for the 
periodic
state of the discharge 
are shown  
in Fig.~\ref{fig_3}a).
Characteristic current peaks corresponding to the breakdown events can be noticed from the results. The temporal evolution of the gap voltage is in correlation with the calculated current showing the fast decrease at the moment of breakdown, which is caused by the charging of the dielectrics. This is illustrated by the representation of the surface charge density $\sigma$ in Fig.~\ref{fig_3}b). 

\begin{figure}[!t]
\centering
\includegraphics[width=80mm]{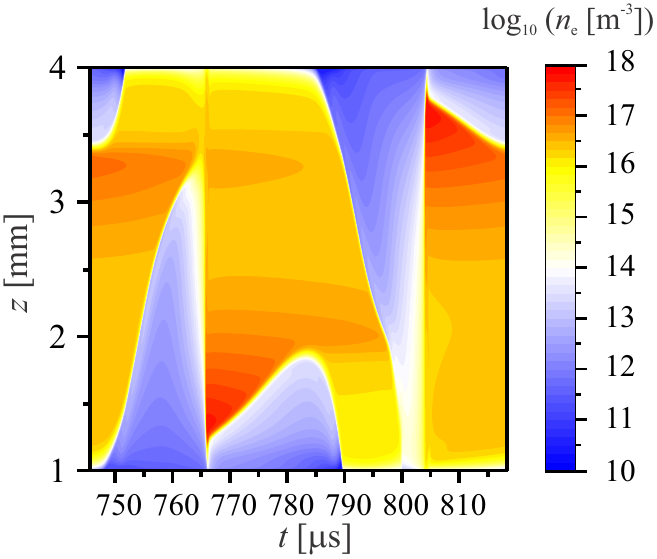}
\caption{Spatiotemporal behaviour of the electron number density during one period for the 
conditions of Figure~\ref{fig_3} (test case 1).}
\label{fig_4}
\end{figure}

The discharge current is mainly determined by the motion of electrons and the maximum number density of electrons is reached at the moment of the highest current. This 
becomes 
obvious from the spatiotemporal evolution of the electron number density $n_\mathrm{e}$ presented in Fig.~\ref{fig_4}. It can be seen that the highest number density of electrons appears at the instant of breakdown in front of the momentary cathode.

\begin{figure}[!t]
\centering
\includegraphics[width=80mm]{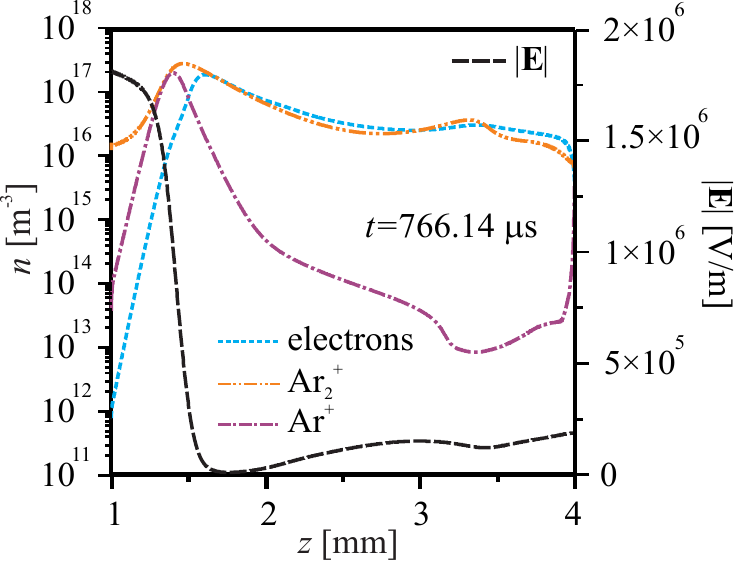}
\caption{Spatial variation of the number densities of electrons, 
	$\mathrm{Ar^+}$ and 
	$\mathrm{Ar_2^+}$ 
as well as of the magnitude of the electric field 
	at
	the moment of maximum current ($t = 766.14$\,\textmu s) for the conditions 
	of Figure~\ref{fig_3} (test case 1).}
\label{fig_5}
\end{figure}

Fig.~\ref{fig_5} shows the spatial distribution of the number densities of molecular ($\mathrm{Ar_2^+}$) and atomic ($\mathrm{Ar^+}$) ions and electrons as well as the magnitude of the electric field at the moment of maximum current in the negative half-period ($t = 766.14$\,\textmu s). Due to strong conversion of $\mathrm{Ar^+}$ in three-body collisions with two argon atoms, 
molecular argon ions are the dominant ionic species at these conditions.  
Electron and ions almost balance each other out in large parts of the discharge region. 
In front of the momentary cathode 
($z = 1$\,mm), 
the number densities of ions are higher than the number density of electrons. 
This leads to the formation of the cathode sheath with high electric field. 
The presented results for this test case are in accordance with results obtained by model calculations for a similar type
of DBD, reported e.g.\ in~\cite{Massines-1998-ID3475}.

\subsubsection{Modelling of a single-filament DBD in argon at atmospheric pressure}
\begin{figure}[!htb]
\centering
\includegraphics[width=80mm]{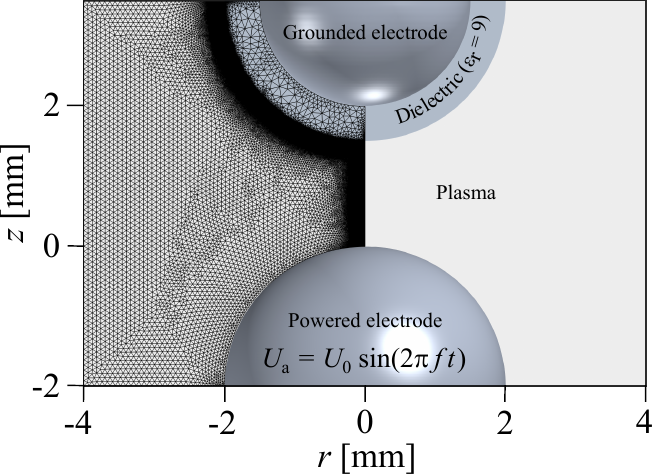}
\caption{Electrode setup of the asymmetric DBD configuration, along with the illustration of the mesh used in test case 2.}
\label{fig_6}
\end{figure}

The second test case deals with time-dependent and spatially two-dimensional (2D) 
modelling 
of a single-filament 
atmospheric-pressure 
DBD in argon. The asymmetric configuration is shown in Fig.~\ref{fig_6}. 
It consists of two hemispherical electrodes with radius $R = 2$\,mm and gap width $d = 1.5$\,mm, in accordance with the setup described in~\cite{Hoder-2011-ID2684}. Here, the curved electrodes are used to fix one filament at the position with shortest electrode distance. The grounded electrode is covered with a dielectric (alumina, $\varepsilon_\mathrm{r} = 9$, $\gamma = 0.02$, thickness $\Delta = 0.5$\,mm), and the metallic electrode 
(stainless steel, $\gamma = 0.07$) 
is powered by 
the sinusoidal voltage $U_\mathrm{a}(t) = U_0 \sin(2 \pi ft)$.
 The voltage amplitude is set to be $U_0 = 3$\,kV and the frequency is $f = 60$\,kHz. 

The focus of the analysis is on the streamer-induced first breakdown event during the first half-period. In this case, there are no volume and surface charges present in the gap or at the dielectric when the voltage is applied. 
In contrast to test case~1,
a 2D model is required to fully describe the transient discharge dynamics and 
the formation
of the constricted discharge channel occurring in the considered geometry at atmospheric pressure~\cite{Hoder-2011-ID2684}.
Due to the high computational costs, 
the discharge is considered to be axisymmetric 
for simplification 
and the  
model equations are 
solved in cylindrical coordinates.

The solution procedure is similar to that described for test case 1. 
In order to reduce the number of degrees of freedom and consequently the computation time, linear Lagrange elements are used here for all equations. Additionally, manual re-meshing of the domain is done to further reduce memory usage and the calculation time. For the pre-phase, covering the process of volume charge accumulation, a mesh consisting of approximately 60\,000 elements is used, whereas for the streamer phase about 500\,000 elements are required to resolve the steep gradients occurring at the streamer head. 
As shown in Fig.~\ref{fig_6} the mesh is refined near the symmetry axis and the dielectric surface with maximum element size of $4$\,\textmu m near the axis and a mesh size varying between $0.1$ and $1.3$\,\textmu m near the plasma-dielectric interface 
boundary. In the other parts of the domain, the mesh is much coarser. The calculations were carried out on a computer with two Intel\textsuperscript{\textregistered} Xeon E5-2690 CPUs, having in total 128 Gb of RAM. The calculations were carried out using 4 cores and in total lasted approximately one week for the presented time-range.

Finally, it should be noted that the electric current is calculated as an integral of the sum of conduction and displacement current density over the electrode surface using the relation
\begin{equation}
I = \int_{S} \left[ \sum_p q_p\mathbf{\Gamma}_p \cdot \boldsymbol{\nu} + \varepsilon_0 \frac{\mathrm{d} \mathbf{E}}{\mathrm{d}t} \cdot \boldsymbol{\nu} \right] \mathrm{d} S\,,
\end{equation}
where $S$ is the surface area of the bare electrode directly exposed to plasma. 

\begin{figure}[t]
 	\centering
 	\includegraphics[scale = 1]{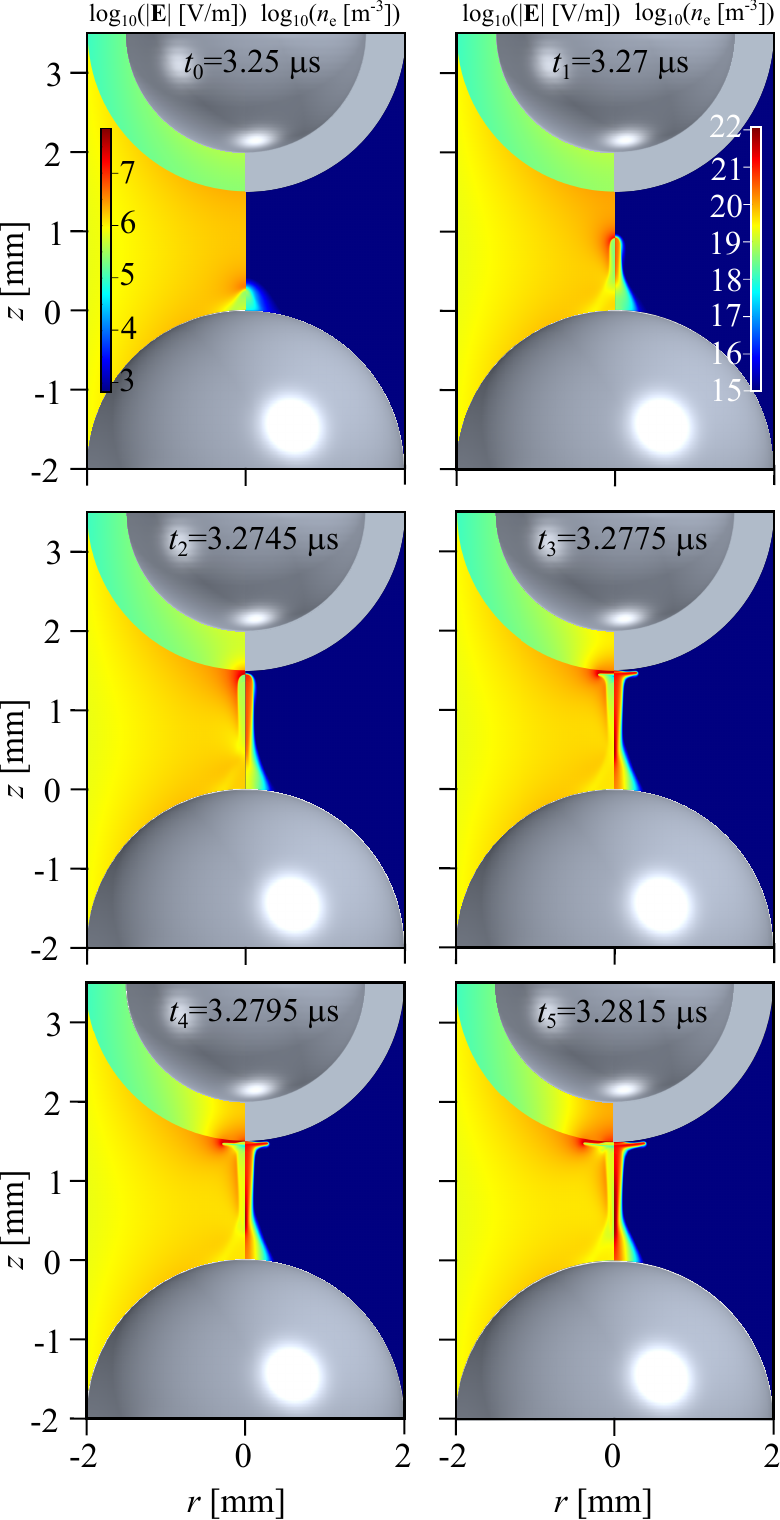}
 	\caption{Spatiotemporal development of the cathode-directed streamer illustrated by the evolution of the electric field magnitude (left-hand-side panel) and electron number density (right-hand-side panel) in characteristic times ($t_0$--$t_5$), noting that the logarithm of these quantities is used for better illustration, and that scales span over the whole time range (test case 2).}
 	\label{fig_7}
\end{figure}

\begin{figure}[t]
\centering
\includegraphics[width=80mm]{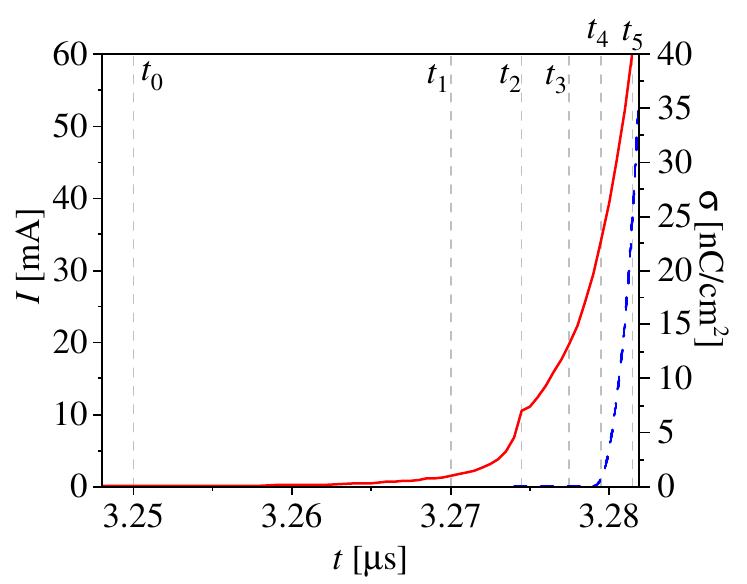}
\caption{Temporal evolution of the electric current 
	$I$	(solid red line) and surface charge density $\sigma$ at the centre of the electrode, i.e.\ at $r = 0$ 
	(dashed blue line) with markers of 
	the characteristic times $t_0$ to $t_5$ corresponding to the spatial profiles presented in Fig.~\ref{fig_7}.}
\label{fig_8}
\end{figure}

In order to illustrate the streamer development, the electron number density and magnitude of the electric field at 
 six characteristic times $t_0$ to $t_5$
 are presented in Fig.~\ref{fig_7}. 
With increase of the applied voltage, Townsend’s phase (pre-phase) commences, during which the accumulation of space charges near the instantaneous anode occurs. 
When the critical density of charged particles is reached (approx. $10^{18}\,\textrm{m}^{-3}$), the streamer phase is initiated (cf. $t_0$ in Fig.~\ref{fig_7}).  
The positive streamer propagation starts in the anode region at $z\approx0.3$\,mm.
As the streamer propagates 
towards the cathode
(cf. $t_1$ in Fig.~\ref{fig_7}) 
a 
steep increase of the electric current $I$ 
can be observed, as displayed in Fig.~\ref{fig_8}. At this point, there are no surface charges accumulated on the dielectric (cf. $t_1$--$t_2$ in Fig.~\ref{fig_8}). Simultaneously, the diameter and magnitude of the electric field increase, reaching its maximum in the vicinity of the cathode (cf. $t_2$ in Fig.~\ref{fig_7}). 
Based on the full-width at half-maximum 
of the radial profile of the electron number density, the streamer diameter  
is determined to be 
$72$\,\textmu m. 
Note that the method described in Nijdam~\textit{et al.}~\cite{Nijdam-2010-ID3010} was used for determination of the streamer radius.

The moment when the streamer reaches the cathode (cf. $t_2$ in Fig.~\ref{fig_7}) manifests as a local maximum (``hump'') in the electric current (cf. Fig.~\ref{fig_8}), as 
also observed in~\cite{Odrobina-1992-ID5473}.
Upon reaching the cathode, the streamer starts propagating along the dielectric surface (cf. $t_3$ to $t_5$ in Fig.~\ref{fig_7}). 
At the same time the surface charge density  starts to increase around $t_4$ (cf. Fig.~\ref{fig_8}). 

\begin{figure}[!t]
	\centering
	\includegraphics[width=2.5in]{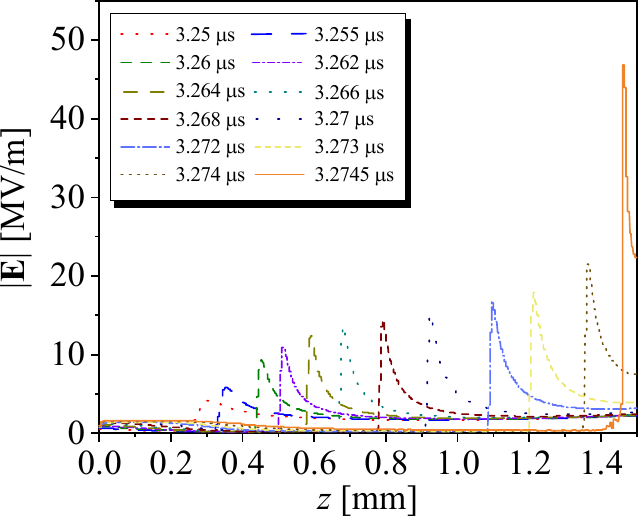}
	\caption{Spatial distribution of the magnitude of the electric field on the symmetry axis for different times $t$ during the volume propagation of the streamer up to $t \approx t_2$.
	}
	\label{fig_9}
\end{figure}

Fig.~\ref{fig_9} shows the spatial variation of the magnitude of the electric field on the symmetry axis for different times. 
A gradual increase of the maximum of $|\mathbf{E}|$ from about 5 to 47\,MV/m is found as the streamer  propagates towards the dielectric surface on the cathode side.

\begin{figure}[!t]
	\centering
	\includegraphics[width=2.5in]{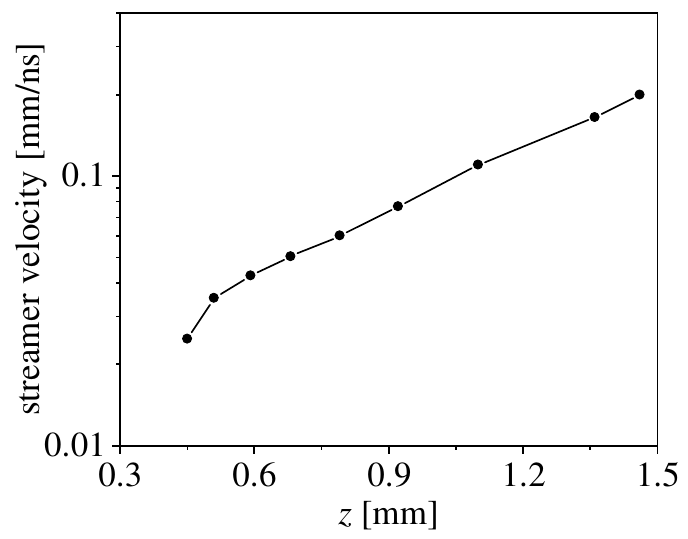}
	\caption{Spatial (i.e. temporal) development of the streamer velocity in the gap during propagation of the streamer head from the anode side (left) towards the cathode (right). The maximum velocity is observed at $t\approx t_2$ in the vicinity of the dielectric boundary on the cathode side at $z=1.5$\,mm. 
}
	\label{fig_10}
\end{figure}

Fig.~\ref{fig_10} illustrates the velocity of the streamer as obtained from the movement of the electric field maximum during  streamer propagation.  
It can be observed that the streamer velocity increases as the streamer propagates in the gap. This is in agreement with 
results of measurements
 obtained for single-filament DBDs in
argon and nitrogen-oxygen gas mixtures for similar geometries~\cite{Kloc-2010-ID2657,Hoeft-2018-ID5099}. A maximum streamer velocity of $0.2\,\textrm{mm/ns}$ is reached in the vicinity of the cathode (cf. $t_2$ in Fig.~\ref{fig_7}), which is in accordance with experimental results for argon DBDs~\cite{Kloc-2010-ID2657} and about one order of magnitude lower than the streamer velocity observed in~\cite{Hoeft-2018-ID5099} for a nitrogen-oxygen DBD.

In comparison to the volume propagation, the velocity of the subsequent streamer propagation along the dielectric surface is 
smaller by one order of magnitude.
The surface streamer propagation continues until a sufficient amount of surface charges is accumulated on the dielectric, leading to the reduction of the electric field and charge carrier production in this region. It is expected that the surface propagation does not occur rotationally uniform but in form of constricted filaments as observed in nitrogen-oxygen DBDs~\cite{Kettlitz-2013-ID3014}. 
Thus, the validity of the present axisymmetric model is limited as soon as the streamer starts to spread along the dielectric surface. 
Further experimental and computational analyses of the surface streamer propagation for the present DBD setup will be part of future investigations.

\section{Summary}
\label{sec:conclusion}
Fluid modelling is a powerful tool for the analysis of the physical and chemical processes of gas discharge plasmas and their applications.
However, 
the management of the equations and source term generation as well as the definition of all required transport and rate coefficients can be a tedious and error-prone process, particularly when lots of species need to be taken into account.

It is shown how the MCPlas toolbox helps to tremendously
speed up the process of model implementation by means of automation. This is achieved by combining functionalities provided by Comsol and Matlab.  
Besides the time-saving aspect, automation virtually eliminates the possibility for errors, which are characteristic for manual implementations of complex reaction kinetic models.
Furthermore, it is discussed how MCPlas supports studies on the influence of different reaction kinetic models and how it facilitates the seamless transition between different model geometries in spatially 1D and 2D configurations.

The practical applicability of the MCPlas toolbox as well as the Comsol model provided by MCPlas is illustrated by two case studies dealing with the analysis of a diffuse and a filamentary DBD in argon.
These examples also demonstrated the advantages of automated code generation by using the same plasma model to study a diffuse discharge in symmetric 1D geometry and a filamentary DBD in asymmetric 2D geometry. 
Implementing the two different models on the basis of the same input files (except for the geometry and discharge parameters) takes only few tens of seconds.

The presented results confirm that even complex processes like the interaction of streamers with dielectric surfaces can be investigated using the models obtained by means of the MCPlas toolbox. An extension of MCPlas to provide the option for spatially three-dimensional geometries is technically possible without further ado. However, the computing time will then be a strongly limiting factor on present computing architectures.

\section*{Acknowledgment}
This work is funded by the Deutsche Forschungsgemeinschaft (DFG, German Research Foundation) - project numbers 368502453 and 407462159.

\ifCLASSOPTIONcaptionsoff
  \newpage
\fi

\bibliographystyle{IEEEtran}

\bibliography{DL_bib}

\begin{IEEEbiography}[{\includegraphics[width=1in,height=1.25in,clip,keepaspectratio]{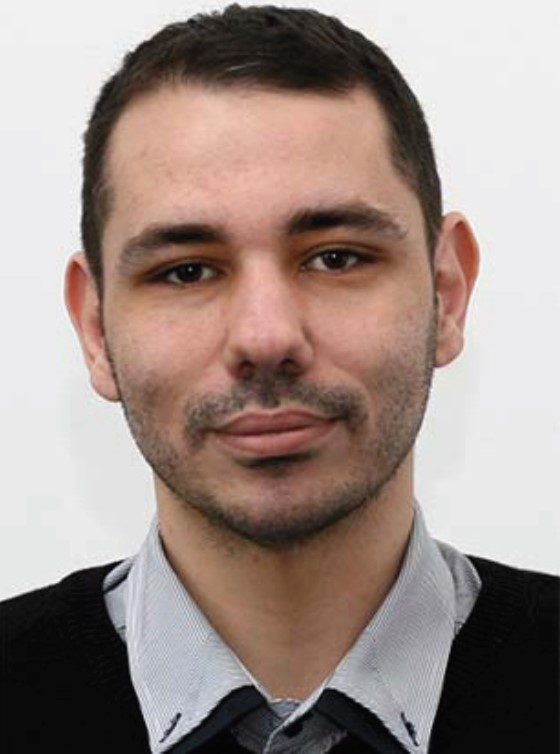}}]{Aleksandar P. Jovanovi\'{c}}
was born in Knjaževac, Serbia in 1984. He graduated in physics in 2010 and received the Ph.D. degree in physics, both from the Faculty of Sciences and Mathematics, University of Niš, Serbia.
From 2011 until 2018, he was employed as a researcher with the Department of Physics at the Faculty of Sciences and Mathematics in Niš. Since 2019, he is researcher with the Plasma Modelling Department, Leibniz Institute for Plasma Science and Technology (INP) in Greifswald, Germany.  His research interest includes plasma modelling with emphasis on electrical discharges (low pressure and DBD), physics of ionised gases, computational physics and statistics.
\end{IEEEbiography}

\begin{IEEEbiography}[{\includegraphics[width=1in,height=1.25in,clip,keepaspectratio]{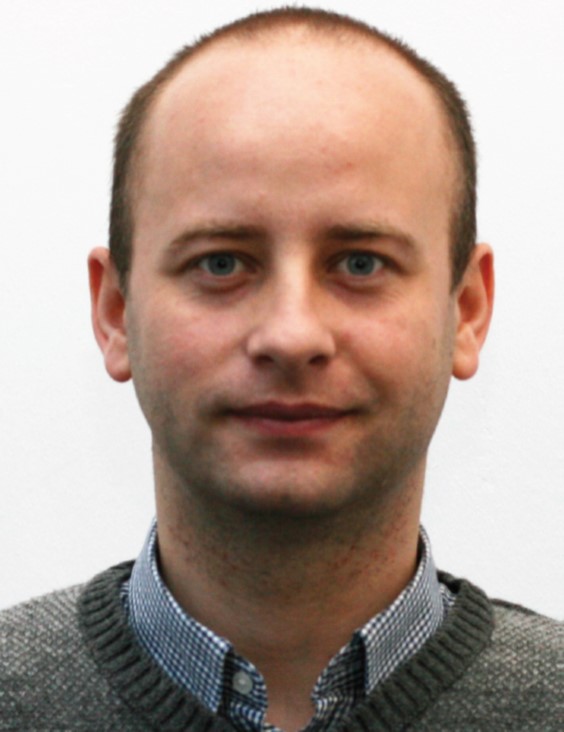}}]{Marjan N. Stankov} 
was born in 1985 in Serbia, where he completed elementary and grammar school. He graduated from the Department of Physics at the Faculty of Sciences and Mathematics in Niš in 2011 and finished his Ph.D. degree in 2015. He was working at the same university until the end of 2017 in the field of low-pressure DC glow discharge plasmas in argon and synthetic air.
Since 2018, he is within Leibniz Institute for Plasma Science and Technology (INP) in Greifswald,
focusing on hydrodynamic modelling of dielectric barrier discharges for plasma-medical application and numerical modelling of microwave discharges suitable for deposition application.
\end{IEEEbiography}

\begin{IEEEbiography}[{\includegraphics[width=1in,height=1.25in,clip,keepaspectratio]{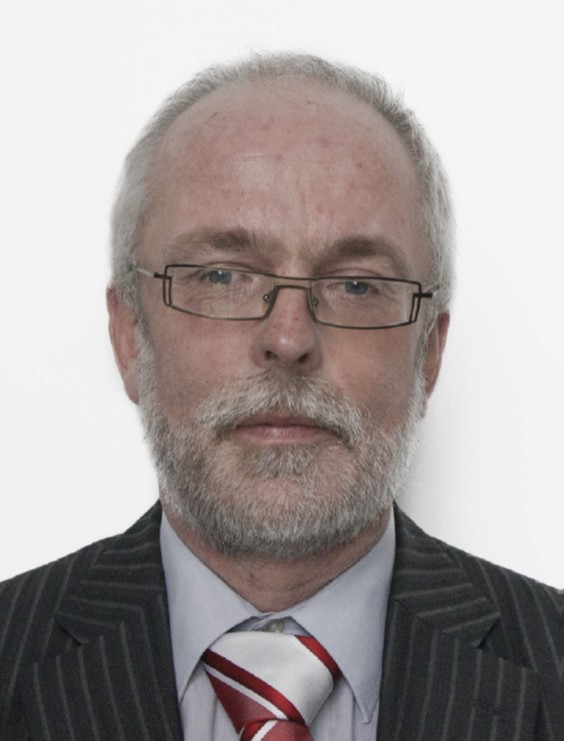}}]{Detlef Loffhagen}
was born in Neustadt am R\"u\-ben\-berge, Germany, on 22 September 1960. He received the Diploma and the Ph.D. degree in physics from University Hannover, Germany, in 1988 and 1992, respectively, and the Dr. rer. nat. habil. degree in theoretical plasma physics from  University of Greifswald, Germany, in 2004.
Between 1993 and 1994, he was with the Institute for  Plasma Physics,  University Hannover. Since May 1994, he has been with the Leibniz Institute for Plasma Science and Technology (INP), Greifswald, Germany, where he became the head of the Department of Plasma Modelling in 2004. His research concerns gas discharge modelling,  plasma chemistry, electron and ion kinetics, and plasma simulation.
\end{IEEEbiography}

\begin{IEEEbiography}[{\includegraphics[width=1in,height=1.25in,clip,keepaspectratio]{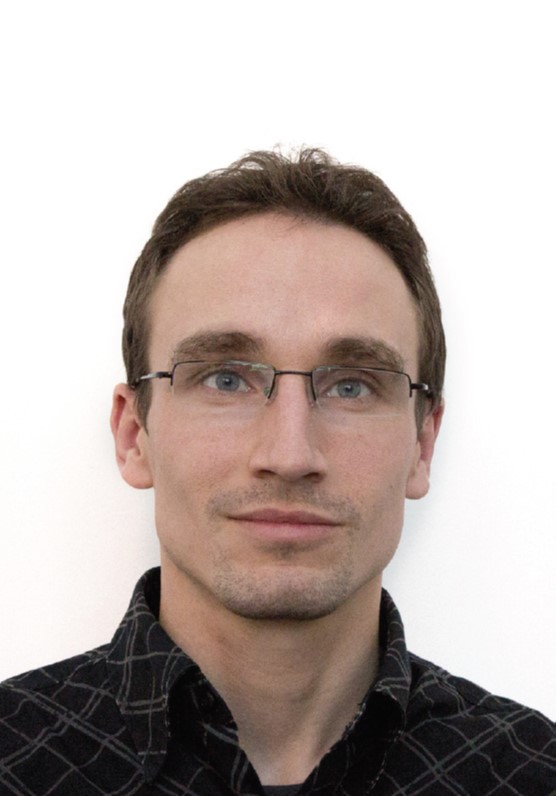}}]{Markus M. Becker} was born in 1982 in Germany. He received the diploma degree in mathematics in 2008 and the Dr. rer. nat. degree in Computational Physics in 2012, both from the University of Greifswald. Since then, he is researcher and project manager at the Leibniz Institute for Plasma Science and Technology (INP) in Greifswald, Germany. His research interests include modelling and applications of dielectric barrier discharges, modelling approaches for non-thermal plasmas, and methods for sustainable data management in plasma science.
\end{IEEEbiography}
\vfill

\end{document}